\documentclass[11pt]{article}
\usepackage[utf8]{inputenc}
\usepackage{amssymb,amsmath}
\usepackage{bm} 
\usepackage{booktabs} 
\usepackage{array}
\usepackage{latexsym}
\usepackage{graphicx}
\usepackage{color}
\usepackage{datetime}
\usepackage[nosort]{cite}
\usepackage{verbatim}
\usepackage{enumerate}
\usepackage{chngpage} 
\usepackage{mathrsfs}
\usepackage{euscript}
\usepackage{psfrag}
\usepackage{subcaption}

\usepackage{datetime}

\usepackage[nosort]{cite}
\usepackage{chngpage} 
\usepackage{setspace}
\usepackage{tensor}
\usepackage{physics}

\usepackage{mciteplus}

\usepackage[colorlinks=true,      linkcolor=blue,      urlcolor=blue,      
            filecolor=blue,      citecolor=blue,       pdfstartview=FitH,     
						pdfpagemode=UseNone,      bookmarksopen=true]{hyperref}  
\usepackage[all]{hypcap}     

\definecolor{cardinal}{rgb}{0.6,0,0}
\definecolor{darkgreen}{rgb}{0,0.4,0}
\definecolor{golden}{rgb}{0.92, 0.7, 0}
\definecolor{midnight}{rgb}{0, 0, 0.5}
\definecolor{darkblue}{rgb}{0, 0, 0.7}
\definecolor{purple}{rgb}{0.5, 0, 0.5}

\def\oneone{\rlap 1\mkern4mu{\rm l}}

\def\coeff#1#2{\relax{\textstyle \frac{#1}{#2}}\displaystyle}

\def\IR{\mathbb{R}}

\def\cL{{\cal L}}

\def\nca{{\zeta}}
\def\ncb{{\xi}}

\def\nBPS#1{$\frac{1}{#1}$-BPS}

\def\mname{maze}

\numberwithin{equation}{section}

\begin{document}

\phantom{AAA}
\vspace{-10mm}

\begin{flushright}
%
%
\end{flushright}

\vspace{1.9cm}

\begin{center}

{\huge {\bf  Waves on Mazes}}\\

{\huge {\bf \vspace*{.25cm}  }}

\vspace{1cm}

{\large{\bf { Iosif Bena$^{1}$,   Rapha\"el Dulac$^{1}$, Anthony Houppe$^{2}$, Dimitrios Toulikas$^{1}$  \\ \vskip  3pt and  Nicholas P. Warner$^{1,3,4}$}}}

\vspace{1cm}

$^1$Institut de Physique Th\'eorique, \\
Universit\'e Paris Saclay, CEA, CNRS,\\
Orme des Merisiers, Gif sur Yvette, 91191 CEDEX, France \\[12pt]

$^2$Institut f\"ur Theoretische Physik, ETH Z\"urich, \\
Wolfgang-Pauli-Strasse 27, 8093 Z\"urich, Switzerland \\[12pt]

\centerline{$^3$Department of Physics and Astronomy}
\centerline{and $^4$Department of Mathematics,}
\centerline{University of Southern California,} 
\centerline{Los Angeles, CA 90089, USA}

\vspace{10mm} 
{\footnotesize\upshape\ttfamily iosif.bena @ ipht.fr,  ahouppe @ phys.ethz.ch, raphael.dulac @ ens.fr, \\
dimitrios.toulikas @ ipht.fr,  warner @ usc.edu} \\

\vspace{1.5cm}
 
\textsc{Abstract}

\end{center}

\noindent One way to describe the entropy of  black holes comes from partitioning momentum charge across fractionated  intersecting brane systems.   Here we construct \nBPS{8} solutions by adding momentum to a maze of M2-brane strips stretched between  M5 branes. Before the addition of momentum, the  \nBPS{4}  supergravity solution describing the maze is governed by a master function obeying a complicated Monge-Amp\`ere equation.  Given such a solution, we show that one can add  momentum waves without    modifying the \nBPS{4} M2-M5 background.  Remarkably, these excitations are fully determined by a layered set of {\em linear} equations.     The fields responsible for carrying the momentum are parameterized by arbitrary functions of a null direction, and have exactly the same  structure as in brane world-volume constructions. The fact that the momentum and flux excitations of the M2-M5-P system are governed by a linear structure brings us one step closer to using supergravity solutions to capture  the entropy of  supersymmetric black-holes.

\begin{adjustwidth}{3mm}{3mm} 

\vspace{-1.2mm}
\noindent

\end{adjustwidth}

\thispagestyle{empty}
\newpage


\tableofcontents

\section{Introduction}
\label{sec:Intro}

\subsection{Overview}
\label{ss:overview}

At the simplest level, this paper gives a construction of \nBPS{8} supergravity solutions in which a momentum wave travels along  a brane intersection.   These solutions are remarkable in their own right, and we show that the BPS equations that govern them reduce to a linear system.  In addition, these solutions also represent a significant advance in the stringy description of black-hole microstructure.  However, we want to set this in context, and so we first describe some recent work  \cite{Bena:2022fzf} that motivated this work and provides a deeper framework for, and understanding of,  such  microstructure.

\subsection{Themelia}
\label{ss:Themelia}

In resolving the microstructure of black holes, one is naturally  led to ask:  what are the fundamental structures in String Theory?  The simplest, and most naive answer is, of course, strings.  However, the answer to this question must be duality invariant.  The  obvious solution is  to include all objects that can be obtained by dualizing strings,  like branes, KK monopoles and brane bound states that preserve sixteen supercharges. 

One can then envision a further extension, to include objects that preserve sixteen supercharges {\it locally}, but preserve only a fraction (or possibly none) of these supercharges when the object is taken as a whole. Such objects were dubbed  {\it themelia} in \cite{Bena:2022fzf}.

A simple example:  A string carrying right-moving momentum is \nBPS{4}, preserving eight supersymmetries.  It is a themelion because, when one ``zooms in'' on the string, one only sees a boosted segment of a string, which preserves 16 supersymmetries.  Another segment of the string also preserves 16 supersymmetries, but  different ones: the supersymmetries depend on the orientation of the string segment \cite{Bena:2011uw}. However, each  set of 16 local supersymmetries  contains a common subset of eight supersymmetries, which make the whole object a \nBPS{4} configuration.

It is natural to expect that themelia will emerge as the fundamental substructure\footnote{The ancient Greek word {\it themelion} ($\theta \epsilon \mu \acute{\epsilon} \lambda \iota o \nu$)   translates to {\it foundation}, or {\it  fundamental structure}.  Unfortunately, the other word for indivisible structure, {\it atomon}, was already taken.} of black hole microstates.   There are three  reasons for this.  First, and most obvious, the themelion is necessarily a bound state because one cannot separate the fundamental charges without breaking some of the 16 local supersymmetries.

Secondly, a system of $N$ identical branes that preserve sixteen supercharges {\it globally} can have an entropy, at most, of order $log N$.   This means that the local structure of a themelion can only account for $log N$ contributions to the entropy.  However, the global structure of a  themelion can involve excitations of its moduli, like shape modes and brane densities, that are parameterized by arbitrary continuous functions, and these can carry an entropy proportional to a power of $N$.  A themelion can only encode such a large entropy in its large-scale, global structure.

The best-studied example is probably the D1-D5 system.  This system carries an entropy of order $\sqrt{N_1 N_5}$.   Taken by itself, it produces a singular black-hole geometry with a Planck-scale horizon.  However, this is not a themelion: it only has eight supercharges locally.  If one adds a KKM and angular momentum, it can be spun out into a supertube, a smooth geometry, with sixteen supercharges locally and eight globally \cite{Mateos:2001qs, Lunin:2001fv, Palmer:2004gu, Lunin:2002iz}.   The degenerate ground states that give rise to the entropy can now be seen as shape modes of the supertube.  Indeed, this example, and the duality-related F1-P system, led to the original fuzzball proposal.

 More broadly, it was observed in \cite{Bena:2022fzf} that all known microstate geometries (and microstate solutions \cite{Bena:2013dka}) are actually based on themelia:  this includes the brane systems underlying three-charge bubbling solutions \cite{Bena:2005va, Berglund:2005vb, Balasubramanian:2006gi},  superstrata \cite{Bena:2011uw, Bena:2015bea} and  the supermaze \cite{Bena:2022wpl}.

The third reason why themelia should be thought of as fundamental constituents of microstate structure, is that a fully back-reacted themelion {\it can never give rise to a classical black hole solution with an event horizon.}   This is because the horizon area, in Planck units, is duality invariant \cite{Horowitz:1993wt}, and so is the same in  {\it all} duality frames. On the other hand, a themelion can always be locally dualized into a stack of $N$ Kaluza-Klein monopoles (KKM's), and this solution is simply empty space with a $Z_N$ orbifold singularity, which is an exact, fully-back-reacted, horizonless string background. 

Independent of geometric considerations, the themelion conjecture states that {\it the fundamental constituents  of black-hole microstructure must be themelia.}  Moreover, because a themelion carries its entropy in its large-scale structure, the supergravity solutions corresponding to coherent collections of themelia should be able to access precisely the degrees of freedom that carry this entropy.  Obviously, supergravity cannot describe string-scale phenomena, but one might hope that supergravity can describe the classical limits of themelia and the degrees of freedom that carry their entropy.  We will refer to this extension  of the themelion conjecture as the {\it geometric themelion conjecture.} 

 The themelion conjecture thus provides an explicit string-theory realization of the fuzzball proposal, while the geometric themelion conjecture provides a precise framework for realizing the goals of the Microstate Geometry programme \cite{Bena:2013dka, Bena:2022rna}.

One can see how the  themelion conjecture can be realized in  the M2-M5-P black hole.   The entropy of this system arises from the fact that each M2 brane can fractionate into $N_5$ strips that can carry momentum independently\footnote{This counting is the M-theory uplift of that of the Type IIA F1-NS5-P black hole \cite{Dijkgraaf:1996cv}.}. Since each of the $N_2 N_5$ strips has four bosonic directions, it is not hard to see that the total entropy (including the fermions) is exactly that of the corresponding black hole, $S = 2 \pi \sqrt{N_2 N_5 N_P}$. Each individual microstate of this system looks like a super-maze of momentum-carrying M2 strips hanging between parallel M5 branes. However, if one considers the brane-brane interactions one finds that the momentum-carrying M2 strips pull on, and deform, the M5 branes. The remarkable feature of this momentum-carrying maze  is that it preserves four supercharges globally, but if one zooms in at any location along the super-maze it preserves locally 16 supercharges. \cite{Bena:2022wpl}.  Thus, the M2-M5-P super-maze is an explicit realization of a themelion that carries all the black hole entropy.

If one could build the supergravity solutions corresponding to this super-maze and show explicitly that these solutions have no horizon, this would establish the geometric themelion conjecture. However, there are several technical hurdles to be overcome. The first is that the \nBPS{4} momentum-less M2-M5 super-maze is governed by a non-linear Monge-Amp\` ere-like equation \cite{Lunin:2007mj,Lunin:2008tf,Bena:2023rzm} and finding cohomogeneity-three solutions (which is the smallest cohomogeneity that gives interesting solutions) is rather involved\footnote{In a recent paper, \cite{Bena:2023rzm}, we have shown in the near-horizon limit of these solutions is related by a change of coordinates to a family of AdS$_3 \times$ S$^3\times$ S$^3 \times \Sigma_2$ solutions, where $\Sigma_2$ is a Riemann surface \cite{Bachas:2013vza}. Since these solutions can be constructed using a linear algorithm, this raises the hope that at least the near-horizon region of certain M2-M5 super-maze geometries will be constructible analytically.}.  The second hurdle is to add momentum to this  M2-M5 super-maze substrate. The latter will be the focus of this paper.

A first step in this direction was achieved in \cite{Bena:2023fjx} using the Born-Infeld action.   One can smear the M2 branes of the super-maze along one of the torus direction, and compactify the solutions to a Type-IIA super-maze consisting of D2 brane strips stretched between parallel D4 branes. The fundamental building block of this IIA super-maze consists of a single D2 brane strip stretched between two parallel D4 branes.  

One can describe such an isolated component entirely in terms of the $SU(2)$ maximally-supersymmetric Yang-Mills theory living on the world-volume of the D4 branes.   This solution is nothing other than the 't~Hooft-Polyakov monopole \cite{Prasad:1975kr,Hashimoto:2003pu,Bena:2023fjx} and this serves as a   \nBPS{4} substrate to which one can add momentum.  Indeed, one can add a null wave in some additional world-volume fluxes. This wave is parameterized by an arbitrary shape function and it does not disturb any of the non-trivial fields of the original 't Hooft-Polyakov monopole \cite{Bena:2023fjx}. Hence, the full \nBPS{8} momentum-carrying solution is constructed in three steps:  first, one builds the non-trivial \nBPS{4} solution to a non-linear set of equations; second, one adds some ``self-dual'' fields that depend on an arbitrary function of the null variable, and lastly, one computes the momentum density of the system, which depends on the square of this arbitrary function.

Such a layered structure is also a feature of all systematic constructions of supersymmetric supergravity solutions with black-hole charges \cite{Bena:2005va, Bena:2004de, Bena:2007kg, Bena:2011dd, Bossard:2012xsa, Bossard:2014ola, Bah:2020pdz, Bah:2021owp, Ceplak:2022wri}. The starting point  is usually a two-charge, \nBPS{4} background.  In five and six dimensions, the most generic solutions start from hyper-K\"ahler, or almost hyper-K\"ahler, four-dimensional spatial base geometry \cite{Gauntlett:2002nw, Gutowski:2004yv, Bena:2005va,Bena:2007kg,Bena:2011dd,Bena:2015bea,Bena:2016ypk,Bena:2017xbt,Tyukov:2018ypq}.  Without imposing additional symmetries, such a base geometry is governed by some very non-trivial, non-linear equation, like the Monge-Amp\` ere system.

The process of adding a third charge in such a manner as to create a themelion,  requires the addition of further dipolar fields, which we think of as  {\it glue}.  The glue is dipolar, and so carries no net charge, but it binds the three fundamental charges together.   As described in \cite{Bena:2011uw} and in \cite{Bena:2022wpl,Bena:2022fzf},  the glue carries precisely the correct {\it local charges} so that, when combined with the global charges,  each and every element of the configuration has 16 supersymmetries locally.    The three overall global charges mean that the themelion is \nBPS{8} but, as we have already noted, the sixteen local supercharges  mean  that one cannot break the configuration apart without breaking the supersymmetry.  It is therefore a bound state and the glue really is glue.

In all these constructions, the BPS equations  governing the addition of the glue and the third charge, which is usually momentum, are linear \cite{Bena:2004de, Bena:2005va,Bena:2007kg,Bena:2011dd, Ceplak:2022wri}.  The only feedback between the overlay of the third charge, its glue and the original, possibly singular,  \nBPS{4} {substrate},  is that the glue smooths out the geometry and fixes some of its moduli.  The  \nBPS{4} solution, and the non-linear equations underlying it, are otherwise unmodified. 

The fact that the same substrate-glue-momentum layered structure appears both in the supergravity description of themelia and in their DBI description \cite{Bena:2023fjx}, leads us to formulate the { ``extended themelion conjecture:''}\\
\indent {\em The addition of momentum on top of a \nBPS{4} themelion substrate can be done using a layered set of linear equations.}

\subsection{Adding momentum to M2-M5 Intersections}
\label{ss:AddMom}

The purpose of this paper is to show that this conjecture correctly describes the addition of momentum to the  M2-M5 super-maze substrate. We use the same type of glue as in the DBI description of \cite{Bena:2023fjx}, and find that, given any  M2-M5 \nBPS{4} solution, one can add a momentum profile specified by an arbitrary function, and obtain a  \nBPS{8} solution by solving {\it a linear system of equations}.  This does not mean that these equations are trivial.  The equation that determines the glue is a homogeneous Laplace equation on a very complicated background.  This solution must then be fed back, quadratically,   as a source for another linear Poisson equation that determines the momentum charge distribution.  Such an  ``upper triangular'' structure of the linear system is  familiar from earlier linear systems governing themelia \cite{Bena:2004de, Bena:2005va,Bena:2007kg,Bena:2011dd, Ceplak:2022wri}.

There is another interesting feature (first observed in \cite{Bena:2022sge}) of our momentum-carrying solutions that greatly simplifies our construction. One can show that even if all the self-dual glue fields depend on a single, arbitrary  function of the null coordinate, this function can be absorbed by a re-definition of the null coordinate so that it only appears in the denominator of a single term in the metric ansatz. Thus, one can construct  a much simpler solution where the arbitrary function is a constant, and then promote the constant to  an arbitrary function. One  can then check explicitly that this is still a  solution.  

We will use the most symmetric \nBPS{4}  super-maze solution as a substrate. This solution has an $SO(4) \times SO(4)$ symmetry and will be reviewed in Section \ref{sec:substrate}.  The first $SO(4)$ corresponds to rotations in the four-dimensional space orthogonal to the M2 branes and the M5 brane, while the second $SO(4)$ corresponds to rotations in the plane of the M5 branes.  

A momentum wave on the M2 brane strips breaks the second $SO(4)$. If the wave is polarized along one of the M5 directions, the symmetry is broken to $SO(3) \times U(1)$ and, upon smearing along the $U(1)$ and reducing to Type IIA String Theory, it becomes a momentum wave on the D4-D2 system.  The construction of this momentum wave in the non-Abelian D4 world-volume theory \cite{Bena:2023fjx}  is reviewed in Section  \ref{sec:Gluing}  and will help us find the glue needed to add momentum while maintaining the themelion structure.  We also show that one can give the wave  a circular polarization that breaks the symmetry  to 
$SU(2) \times U(1)$. We will construct both types of waves in Section \ref{sec:NewSol}.
Section \ref{sec:Conc} contains our final remarks.

\section{Adding momentum to the M2-M5 system - the DBI analysis}
\label{sec:Gluing}

An M2 brane strip stretched between two M5 branes can be smeared along one of the M5 brane directions and reduced to Type IIA String Theory, becoming a D2 strip stretched between two D4 branes. From the perspective of the $SU(2)$ non-Abelian D4-brane world-volume theory this solution is nothing other than a 't Hooft-Polyakov monopole solution in 3+1 dimensions \cite{Prasad:1975kr} that is independent of the fourth D4 world-volume direction  \cite{Hashimoto:2003pu,Bena:2023fjx}. This is the common D2-D4 direction, along which one can add a null momentum wave involving several non-trivial fields in the D4 world-volume theory  \cite{Bena:2023fjx}. A simpler solution is a semi-infinite D2 ending on a D4 brane, to which one can also add a momentum wave using the same D4 world-volume fields. In this Section we review this construction and use it to reveal the fields that one must use in supergravity to add momentum to this brane system.

It is well known that a semi-infinite F1 or D1 string ending on a D3 brane pulls it and forms a spike \cite{Callan:1997kz, Constable:1999ac}, and this can be described as an Abelian monopole in the Born-Infeld action. By T-duality, one can see that the same gauge configuration also describes a semi-infinite D2 furrow  bending a D4 brane.  This solution has sixteen supercharges locally\footnote{As do all  solutions of the Abelian DBI action.} and eight globally,  and so it is a \nBPS{4} themelion. The furrow is extended along the D2-D4 common direction, $y$, and looks like a spike in the  $\IR^3$ spanned by the other three directions of the D4-brane world-volume.   

Since the solution is spherically symmetric in this $\IR^3$ we use coordinates, $(u, \theta, \phi)$ where $u$ is the radius and  $(\theta, \phi)$ are angles on the $S^2$.  We  define  $\hat{\theta}, \hat{\phi}$ to be  flat indices for frames on the  $S^2$.   The D4 scalars and Maxwell fields that describe the D2 spike take the form:
\begin{equation}
\Phi~=~ \frac{b}{u}  \,,   \qquad F_{\hat{\theta} \hat{\phi}}~=~\partial_u\Phi  \,. 
\label{BIon}
\end{equation}
The scalar $\Phi$ determines the displacement of the D4-brane in the  D2-brane direction, $z$, orthogonal to the D4 brane. The non-trivial profile of $\Phi$ describes the spike.  This profile  sources the monopole configuration of the Maxwell field, and the result is a solution to the DBI equations.  Note that this solution is independent of the common D2-D4 direction, $y$.

%
%

One can add momentum to the D2-D4 brane solution by turning on a magnetic and an electric field in the D4 world-volume theory:
\begin{equation}
    F_{uy}=-F_{u0}=\frac{f(y-t)}{u^2}\,,
    \label{DBIglue}
\end{equation}
with $f(y-t)$ an arbitrary function. One can check that adding these fields does not disturb the fields already present in the original D4-D2 solution \eqref{BIon}. The non-trivial electric and magnetic fields generate a Poynting vector, giving rise to a  momentum density along $y$, and a net global momentum charge.  

Thus the global charges of this solution are:
\begin{equation}
Q^{\rm D4}_{\mathrm{u}\theta\phi\mathrm{y}}  \,, \qquad D^{\rm D2}_{\mathrm{zy}} \,, \qquad 
Q^{\rm P}_{\mathrm{y}} \,,
\end{equation}
where the subscripts denote the directions.

 \begin{figure}[h!]
	\begin{subfigure}[h!]{0.35\linewidth}
		\includegraphics[width=\linewidth]{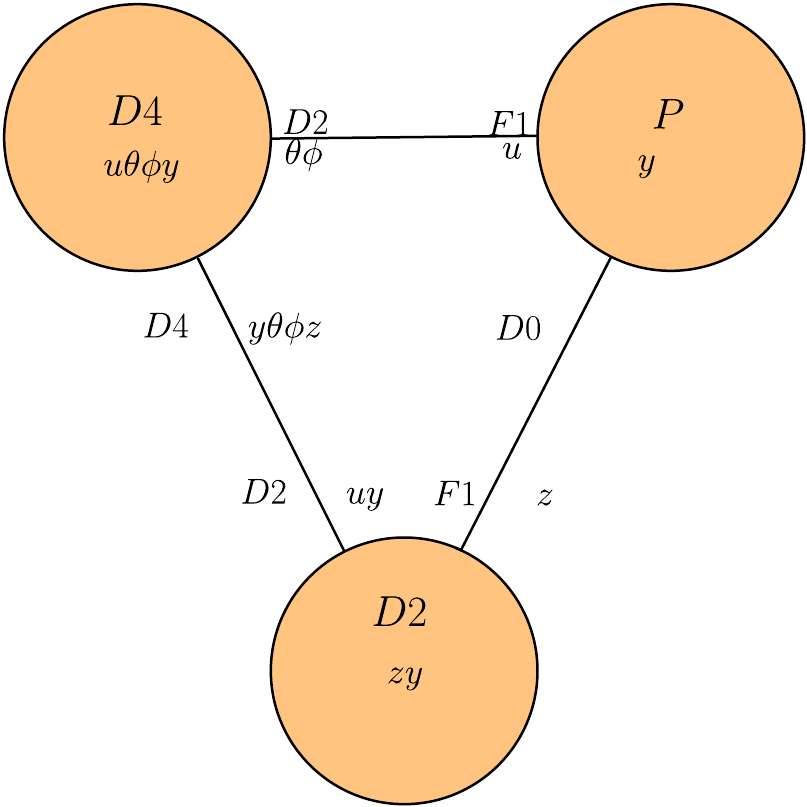}
		\caption{Type-IIA charges and glue}
	\end{subfigure}
	\hfill 
	\begin{subfigure}[h!]{0.35\linewidth}
		\includegraphics[width=\linewidth]{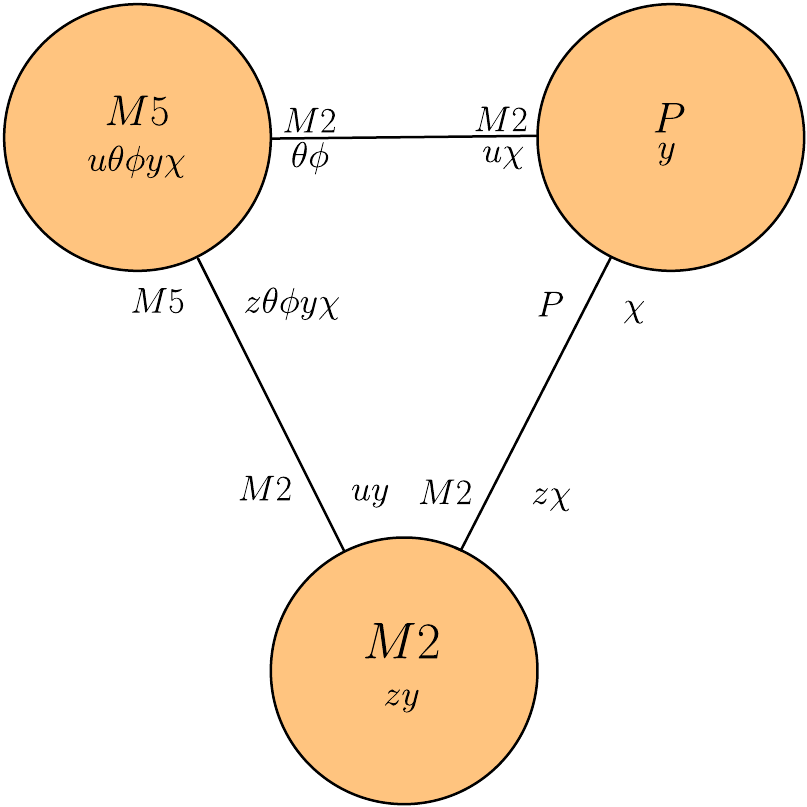}
		\caption{M theory charges and glue.}
	\end{subfigure}%
	\caption{Global and local charges of our solution in Type IIA String Theory (a) and M theory (b). The brane charges in the circles are global charges. The branes on the branches are the ``glue'' one needs to add in order to have enhance the local supersymmetry to 16 supercharge. The two ``glue'' branes on a given branch have the same local charge.}
	\label{gluing}
\end{figure}

The brane profile also sources a collection of dipole charges, the glue:   
\begin{equation}
		Q^{\rm D4}_{\mathrm{z}\theta\phi\mathrm{y}}\,, \quad Q^{\rm D2}_{\mathrm{uy}}\,, \quad
		Q^{\rm D2}_{\theta\phi}\,, \quad  Q^{\rm F1}_{ \mathrm{u}}\,, \quad
		Q^{\rm D0} \,, \quad Q^{\rm F1}_{ \mathrm{z}}\,.
\end{equation}
It is easy to see the intuitive origin of these dipoles. The first two reflect the bending of the D4 brane pulled on by the D2 branes, and are also present in the absence of a momentum wave. They correspond to the left-hand side of the triangle in Figure \ref{gluing}(a). The remaining dipole charges are linear  in the profile function and thus create no net charge. The $Q^{\rm D2}_{\theta\phi}$ charge comes from $F_{uy}$, which sources a $C_{0\theta\phi}$ through the WZ term. Similarly, $Q^{\rm D0}$ is sourced by $F_{uy}\wedge F_{\theta \phi}$ in the WZ term. The remaining F1 charge arises through the DBI action. Both are sourced by $F_{0u}$: $B_{0u}$ couples directly, while $B_{0z}$ couples via the pull-back $\partial_u \Phi$.  The complete set of global and dipole charges are depicted in Figure \ref{gluing}, along with their M-theory uplifts.

A more precise analysis \cite{Bena:2023fjx} enables one to read off the local charges of this solution directly from the $\kappa-$symmetry of the DBI action.  The  simplest way to express all the charges  is to introduce two angles
\begin{align} \label{wingling angle}
    &\tan\alpha \equiv \partial_u\Phi \,,\\
    &\tan \beta \equiv \frac{F_{yu}}{\sqrt{1+(\partial_u\Phi)^2}}\,.
\end{align}
The angle $\alpha$ can be thought as the local slope of the D2-D4 spike, which parameterizes how much the D2 pulls on the D4 world-volume. The angle $\beta$ can be thought as the local pitch of the momentum-carrying wave. The three charge densities that contribute to the global charges of the solution can be written as:
\begin{align}
	&Q^{\rm D4}_{\mathrm{u}\theta\phi\mathrm{y}}= M \cos^2\!\alpha\cos^2\!\beta \,, \\
	&D^{\rm D2}_{\mathrm{zy}}=M \sin^2\!\alpha\cos^2\!\beta \,,\\
	&Q^{\rm P}_{\mathrm{y}}=M \sin^2\!\beta \,.
\end{align}

The six local charges, that form the glue needed to give the themelion sixteen-local-supercharges structure to the solution are:
\begin{eqnarray}
		Q^{\rm D4}_{\mathrm{z}\theta\phi\mathrm{y}}=M \cos^2\! \beta \cos \alpha \sin \alpha \,, & Q^{\rm D2}_{\mathrm{uy}}=-M \cos^2\! \beta  \cos \alpha \sin \alpha \,, \\
		Q^{\rm D2}_{\theta\phi}=M \cos\beta \sin\beta\cos\alpha \,, &Q^{\rm F1}_{ \mathrm{u}}=- M \cos\beta\sin\beta\cos\alpha \,, \label{M2-glue}\\
		Q^{\rm D0} =M \cos\beta \sin\beta\sin\alpha \,, &Q^{\rm F1}_{ \mathrm{z}}=- M \cos\beta\sin\beta\sin\alpha\,.
\end{eqnarray}

The uplift of this configuration to M-theory is a momentum wave on an M2 strip meeting an M5.  This is depicted on the right-hand side of Figure \ref{gluing}, where $\chi$ is the M theory direction.  Since the basic configuration is intersecting M2 and M5 branes carrying momentum we will think of this as a kind of super-maze  \cite{Bena:2022wpl}, except that the glue of this system is not the same\footnote{In \cite{Bena:2022wpl}, the glue on the M5-M2 and the M2-P branch is the same, but the glue on the M5-P branch corresponds to M5 and P rather than two species of M2 branes, as depicted along the top of Figure \ref{gluing}(b).}  as that of  \cite{Bena:2022wpl}.

\section{The M2-M5 substrate}
\label{sec:substrate}
We review the supergravity solutions for M2-M5 intersections \cite{Lunin:2007mj,Lunin:2008tf,Bena:2023rzm}, to which we will add momentum in the next section. Our discussion here closely parallels that of \cite{Bena:2023rzm}. 

\subsection{The brane configuration}
\label{ss:config}

The M2 branes will be taken to lie along the 
$(x^0, x^1, x^2) =(t, y, z)$ directions, and they have the $(x^0, x^1) =(t, y)$ directions in common with the M5-branes. These branes will extend along the $(x^0, x^1,x^3,x^4,x^5,x^6)=(t,y, \vec u)$ directions. The spatial directions transverse to the branes will be denoted by $(x^7,x^8,x^9,x^{10})= \vec v$.  

These configurations have  eight supercharges, and are thus  \nBPS{4}  supergravity solutions, and their supersymmetries satisfy the projection conditions
\begin{equation}
  \Gamma^{012} \, \varepsilon  ~=~ - \varepsilon \,,   \qquad  \Gamma^{013456} \, \varepsilon  ~=~\varepsilon \,.
 \label{projs1}
\end{equation}
The indices on the gamma matrices are frame indices taken along the  directions of the M2's and M5's. 
Recalling that in eleven dimensions one has $ \Gamma^{0123456789\,10} =  \oneone$, one sees that (\ref{projs1})  implies
\begin{equation}
 \Gamma^{01789\,10} \, \varepsilon  ~=~-\varepsilon \,,
 \label{projs2}
\end{equation}
and hence one can add another set of M5 branes along the directions $01789\,10$ without breaking supersymmetry any further.  We will denote this second possible set of branes by M5', but we will not actively include sources for such branes.

The \nBPS{4} solutions of interest have an eleven-dimensional metric of the form:
\begin{equation}
\begin{aligned}
ds_{11}^2 ~=~  e^{2  A_0}\, \Big[ -dt^2 &~+~ dy^2 ~+~ e^{-3  A_0} \, (-\partial_z w )^{-\frac{1}{2}}\, d \vec u \cdot d \vec u  ~+~ e^{-3  A_0} \, (-\partial_z w )^{\frac{1}{2}}\, d \vec v \cdot d \vec v \,    \\
  &  ~+~  (-\partial_z w ) \, \big( dz ~+~(\partial_z w )^{-1}\,   (\vec \nabla_{\vec u} \, w )  \cdot  d \vec u \big)^2  \Big]\,.
\end{aligned}
 \label{11metric}
\end{equation}
This metric is conformally flat along time and the common M2-M5 direction, $(t,y) \in \IR^{(1,1)}$, and also along the  M5  directions, parameterized by  $\vec u \in \IR^4$,  and on a  transverse $\IR^4$ parameterized by $\vec v \in \IR^4$.    The metric involves a non-trivial fibration of the ``M-theory direction,'' $z$, over this internal  $\IR^4$.    The  constraints on, and relationships between, the functions $A_0(\vec u, \vec v, z)$ and $w(\vec u, \vec v, z)$ will be discussed below, and, for obvious reasons, we require $\partial_z w <0$.

We will use the set of frames:
\begin{equation}
\begin{aligned}
e^0 ~=~  & e^{A_0}\, dt \,, \qquad e^1~=~  e^{A_0}\, dy \,,  \qquad e^2 ~=~   e^{A_0} (-\partial_z w )^{\frac{1}{2}} \, \Big( dz ~+~(\partial_z w )^{-1}\,  \big (\vec \nabla_{\vec u} \, w \big)  \cdot  d \vec u \Big) \,, \\
e^{i+2}~=~  & e^{- \frac{1}{2} A_0} \, (-\partial_z w )^{-\frac{1}{4}}\,  du_i \,, \qquad e^{i+6} ~=~  e^{- \frac{1}{2} A_0} \, (-\partial_z w )^{\frac{1}{4}}\,  dv_i\,,   \qquad {i = 1,2,3,4}   \,.
\end{aligned}
 \label{11frames}
\end{equation}
The three-form vector potential is given by:
\begin{equation}
C^{(3)} ~=~   - e^0 \wedge e^1 \wedge e^2 ~+~ \frac{1}{3!}\, \epsilon_{ijk\ell} \,  \big((\partial_z w )^{-1}\, (\partial_{u_\ell} w) \,  du^i \wedge du^j \wedge du^k ~-~ (\partial_{v_\ell} w)  \, dv^i \wedge dv^j \wedge dv^k  \big)  \,,
 \label{C3gen}
\end{equation}
where $\epsilon_{ijk\ell}$ is the $\epsilon$-symbol on $\IR^4$. 

\subsection{The \mname\ function}
\label{ss:master}

Define the following combinations of the functions that determine the metric and fluxes:
\begin{equation}
F_1   ~\equiv~ (-\partial_z  w)^{\frac{1}{2}} \,e^{-3   A_0} \,, \qquad  F_2  ~\equiv~ (-\partial_z  w)^{-\frac{1}{2}} \,e^{-3   A_0}~-~  (\partial_z w )^{-1} \, (\nabla_{\vec u} \,w )\cdot  (\nabla_{\vec u} \, w )\,,
 \label{F12defns}
\end{equation}
and denote the Laplacians on each $\IR^4$ via:
\begin{equation}
{\cal L}_{\vec u} ~\equiv~  \nabla_{\vec u} \cdot \nabla_{\vec u} \,, \qquad {\cal L}_{\vec v} ~\equiv~  \nabla_{\vec v} \cdot \nabla_{\vec v} \,.
 \label{Laps}
\end{equation}

One then finds that the supersymmetry variations lead to the BPS equations: 
\begin{equation}
\cL_{\vec v}  w  ~-~  \partial_z F_1    ~=~ 0\,, \qquad \cL_{\vec u}  w  ~+~  \partial_z F_2~=~ 0  \,.
 \label{BPS1}
\end{equation}
The equations of motion also require:  
\begin{equation}
\cL_{\vec v}  F_2 ~+~  \cL_{\vec u}   F_1    ~=~ 0\,.
 \label{BPS2}
\end{equation}
One should note that this is essentially an integrated form of (\ref{BPS1}): if one takes $\partial_z$ of  (\ref{BPS2}), one obtains an identity that is a trivial consequence of  (\ref{BPS1}).

This system can be solved by introducing a pre-potential, $G_0$, and setting:
\begin{equation}
w ~\equiv~ \partial_z G_0  \,, \qquad F_1 \equiv e^{-3  A_0} \, (-\partial_z w )^{\frac{1}{2}} ~=~ {\cal L}_{\vec v}  G_0 \,.
 \label{solfns}
\end{equation}
The remaining equations are then equivalent to: 
\begin{equation}
 F_2 \equiv e^{-3  A_0} \, (\partial_z w )^{-\frac{1}{2}} ~-~ (\partial_z w )^{-1} \, (\nabla_{\vec u} \,w )\cdot  (\nabla_{\vec u} \, w ) ~=~  - {\cal L}_{\vec u}  G_0  \,.
 \label{reln1}
\end{equation}
Using  (\ref{solfns}) in this equation to eliminate $w$ and $A_0$ yields the Monge-Amp\`ere-like equation for the pre-potential:
\begin{equation}
 {\cal L}_{\vec v} G_0 ~=~  ({\cal L}_{\vec  u} G_0)\,(\partial_z^2 G_0) ~-~ ( \nabla_{\vec u} \partial_z G_0)\cdot  (  \nabla_{\vec u} \partial_z G_0)\,.
 \label{maze-eq}
\end{equation}
Given a solution to (\ref{maze-eq}) for the \mname\ function, $G_0$, one can obtain all the metric/flux functions from (\ref{solfns}).

While this non-linear equation is daunting, it has been shown that solutions exist, and can be constructed in an iterative expansion \cite{Lunin:2007mj,Lunin:2008tf}.  Moreover, by imposing spherical symmetry and taking a near-brane limit, one can reduce that maze function to two variables, and it can be re-cast as a linear system \cite{Lunin:2007ab,DHoker:2008lup,DHoker:2008rje,DHoker:2008wvd,DHoker:2009lky, DHoker:2009wlx,Bobev:2013yra,Bachas:2013vza,Bena:2023rzm}.  We will therefore take this solution as  ``given,'' and assume that we have some form of intersecting M2-M5 substrate on which we will erect momentum modes.

\section{The \nBPS{8}  M2-M5-P themelion}
\label{sec:NewSol}

\subsection{The Ansatz for the metric and flux}
\label{ss:metricfluxAnsatz}

We will impose an $SO(4)$ symmetry transverse to the branes, and  write the metric in the directions, $(x^7, x^8,x^9,x^{10})$, in  terms of a radial coordinate $v$, and an $S^3$, with unit metric $d{\Omega'}_3^2$. We will not, {\it a priori}, assume that the metric in these directions is conformal to the $\IR^4$ metric (\ref{4metrics-1a}).

We are going to consider two possible structures along the $3456$ directions of the M5 brane:
\newpage

\begin{itemize}
\setlength\itemsep{-2pt }
\item[]  {\bf Choice (i):}
\item[] We  take these directions to have an $SO(3) \times \IR$ symmetry, in which $x^6 = \chi $ is  $\IR$, or $S^1$, and the remaining directions are described by a radial coordinate, $u$, and  spheres, $S^2$, with unit metric  $d{\Omega}_2^2$.  We allow the scale factors in front of  $du^2$, $d{\Omega}_2^2$ and $d\chi^2$ to be independent arbitrary functions of $(u,v,z)$.   The polarization density of the null wave will be directed along $x^6=\chi$. This solution can be compactified to Type IIA String Theory along $\chi$ to give the supergravity solution corresponding to the D2-D4-P configuration in Section \ref{sec:Gluing}.
\end{itemize} 
\begin{itemize}
\setlength\itemsep{-2pt }
\item[]  {\bf Choice (ii):}
\item[] We impose an $SU(2)_L \times U(1)$ symmetry by introducing a radial coordinate, $u$, and (possibly {\it squashed}) spheres, $S^3$, whose  metric we take to be that of a Hopf fibration over $S^2$.  We allow the scale factors in front of  $du^2$, the $S^2$ base of the fibration, and the Hopf fiber to be independent arbitrary functions of $(u,v,z)$. The  polarization density of the null wave will be directed along the Hopf fiber.
\end{itemize} 
Finally, following \cite{Lunin:2007ab,Bena:2023rzm} and, as in (\ref{11metric}), we are also going to allow a non-trivial fibration of the $z$-direction over $du$.

To be more specific, we are going to analyze two possible  geometries on the M5 branes:
\begin{equation}
{\rm Choice~(i)}: ~ds_{4}^2  ~\equiv~     du^2  ~+~ u^2 \, d\Omega_2^2 ~+~ d\chi^2 
\,, \quad {\rm or} \quad  {\rm Choice~(ii)}: ds_{4}^2 ~\equiv~     du^2  ~+~ u^2 \, d\Omega_3^2 \,,
 \label{4metrics-1a}
\end{equation}
and the metric on the $\IR^4$ orthogonal to the M5 branes
\begin{equation}
 {ds'}_4^2  ~\equiv~  dv^2  ~+~ v^2 \,  d\Omega'{}_3^2    \,.
 \label{4metrics-2a}
\end{equation}
Here, $d\Omega_n^2$ is the maximally symmetric metric on a unit-radius $S^n$, and $\chi$ is some flat direction of which the solution is independent.
The corresponding Laplacians are:
\begin{equation}
 {\cal L}_{u} G ~=~  \frac{1}{u^n} \partial_u \big(u^n \partial_u G \big) \,, \qquad  {\cal L}_{v} G ~=~  \frac{1}{v^3} \partial_v \big( v^3 \partial_v G\big)\,,
 \label{sphLaps}
\end{equation}
where $n=2$ for Choice (i) and $n=3$ for Choice (ii).

We want to construct a solution that has the charges of the super-maze constructed in Section \ref{sec:Gluing}, and not the charges of the super-maze of \cite{Bena:2022wpl}, so the momentum of the solution will only be along the common M2-M5 direction, $y$ (and not along the pure M2-direction $z$).   We therefore introduce null coordinates, $\nca,\ncb$, and take  $(x^0+ x^1, x^1\! -\!x^0,x^2) =(\nca, \ncb , z)$, and assume the wave is a function of $\xi$.  We also expect that the null wave profile can be an arbitrary function, $F(\ncb)$, and we  will show  that this expectation is indeed correct. 

We therefore use the metric Ansatz :
\begin{equation}
\begin{aligned}
ds_{11}^2 ~=~  e^{2  A_0}\, \bigg[   d \ncb \, \bigg( P \,  d  \ncb    ~+~ 2\, \bigg( \frac{ d \nca}{F(\ncb) } ~+~ k \, \hat \sigma_3  \bigg)\bigg) &  ~+~ e^{2 A_1} \,  ds_4^2   ~+~ e^{2 A_2} \,  {ds'}_4^2
 \\ &   ~+~ e^{2 A_3} \, \Big( dz ~+~B_1 \, du \Big)^2  \bigg]\,,
\end{aligned}
 \label{11metric-null}
\end{equation}
where  the four-dimensional metrics are given by:
\begin{equation}
 ds_{4}^2 ~\equiv~     du^2  ~+~ \coeff{1}{4} \, u^2 \, e^{2 A_4}\,(\sigma_1^2 +\sigma_2^2\big) ~+~ \coeff{1}{4} \,  u^2 \,e^{2 A_5}\,\sigma_3^2 \,,
 \label{4metrics-1b}
\end{equation}
and
\begin{equation}
 {ds'}_4^2  ~\equiv~  dv^2  ~+~ \coeff{1}{4} \, e^{2 A_6}\,  v^2 \, \,({\sigma'}_1^2 +{\sigma'}_2^2+{\sigma'}_3^2\big)    \,.
 \label{4metrics-2b}
\end{equation}
For Choice (ii), we take the $\sigma_i$ to be  the left-invariant $1$-forms on $S^3$:
\begin{equation}
\begin{aligned}
\sigma_1 ~=~  &   \cos \varphi_3 \, d \varphi_1 ~+~ \sin \varphi_3 \sin \varphi_1\, d \varphi_2  \,, \\ 
\sigma_2 ~=~  &   \sin \varphi_3 \, d \varphi_1 ~-~ \cos \varphi_3 \sin \varphi_1\, d \varphi_2  \,, \\ 
\sigma_3 ~=~  &    d \varphi_3 ~+~ \cos  \varphi_1\, d \varphi_2  \,,
\end{aligned}
 \label{1forms}
\end{equation}
with the similar expressions for the ${\sigma'}_i$, but with $\varphi_j \to {\varphi'}_j$.   The polarization vector, $\hat \sigma_3$ is set equal to $\sigma_3$, pointing along the Hopf fiber:
\begin{equation}
\hat \sigma_3 ~=~\sigma_3\,.
 \label{sighat1}
\end{equation}

For Choice (i), the $\sigma_i$ are:
\begin{equation}
\sigma_1 ~=~    2\,  d \theta  \,, \qquad 
\sigma_2 ~=~   2\,  \sin \theta \, d \phi \,,   \qquad 
\sigma_3 ~=~    \frac{2}{u} \,  d \chi\,,
 \label{alt1forms}
\end{equation}
so that the metric  (\ref{4metrics-1b}) is  becomes 
\begin{equation}
ds_{4}^2 ~=~  du^2  ~+~ u^2 \,  e^{2 A_4}\, \big(d\theta^2  + \sin^2\theta\, d\phi^2 \big)  ~+~ e^{2 A_5}\, d\chi^2 \,. 
 \label{4metrics-1c} 
 \end{equation}
The polarization vector, $\hat \sigma_3$, now points along $d\chi$:
\begin{equation}
\hat \sigma_3 ~=~d\chi \,.
 \label{sighat2}
\end{equation}
The metric ${ds'}_4^2$ remains the same for both choices, with ${\sigma'}_i$ being the left-invariant $1$-forms on ${S'}^3$.  %

In this Ansatz, the functions $P, k$ and $A_n$, $n=0,1,\dots, 6$ are, as yet arbitrary functions of $(u,v,z)$.  The only dependence on $\ncb$ appears through the single function, $F(\ncb)$ in the metric.

We use the orthonormal frames:
\begin{equation}
\begin{aligned}
e^0 ~=~  & \frac{e^{A_0} }{\sqrt{P} }\,\bigg( \frac{ d \nca}{F(\ncb) } + k \, \hat \sigma_3  \bigg)  \,, \qquad e^1~=~    \frac{e^{A_0} }{\sqrt{P} }\, \bigg(P \,  d  \ncb    +   \frac{ d \nca}{F(\ncb) } + k \, \hat \sigma_3 \bigg)   \,, \\ e^2 ~=~ &  e^{A_0 +A_3} \, \Big( dz  + B_1 \, du \Big)  \,,  \qquad e^3 ~=~  e^{A_0 +A_1} \, du  \,,   \qquad e^4 ~=~  e^{A_0 +A_2} \, dv  \,, \\
e^{5,6}~=~  & \coeff{1}{2} \, u \,  e^{A_0 +A_1 +A_4} \, \sigma_{1,2} \,, \qquad e^{7}~=~    \coeff{1}{2} \, u \,  e^{A_0 +A_1 +A_5} \, \sigma_{3} \,, \qquad   e^{8,9,10} ~=~  \coeff{1}{2} \, v \,  e^{A_0 +A_2 +A_6} \, {\sigma'}_{1,2,3}       \,.
\end{aligned}
 \label{11frames-null}
\end{equation}

Rather than making a general Ansatz for the potential, $C^{(3)}$, we find it easier to make the most general possible Ansatz for the fluxes, $F^{(4)}$,  in a manner that is consistent with all of the symmetries.  The frames $e^5$ and $e^6$ must appear as $e^5 \wedge e^6$, and the frames $e^8, e^9$ and $e^{10}$ must appear as $e^8 \wedge e^9\wedge e^{10}$, while the remaining six frames, $e^0, e^1,e^2,e^3,e^4$ and $e^7$,  can appear in any combination.  This means that there are, in principle, $36$ functions that can appear when $F^{(4)}$  is expanded in frames.  We thus introduce  $36$ functions of $(u,v,z)$  into our Ansatz.

One could use the general properties of null waves to simplify the Ansatz for $F^{(4)}$, but this turns out to be unnecessary.  We will, however, note that in addition to the fluxes sourced by the background M2 and M5 branes, we expect momentum waves to involve flux components with legs along $d\ncb$ and not $d\nca$.  In terms of frames, this means that the momentum waves source flux components involving only $e^1 \! - \!  e^0$.  This is indeed what we find from solving the BPS equations.

\subsection{The supersymmetries}
\label{ss:susies}

The four supersymmetries of the \nBPS{8} system are defined by adding a momentum projector to the M2 and M5 brane projectors in (\ref{projs1}): 
\begin{equation}
 \Gamma^{01} \, \varepsilon  ~=~ - \varepsilon \,,   \qquad   \Gamma^{012} \, \varepsilon  ~=~ - \varepsilon \,,   \qquad  \Gamma^{013456} \, \varepsilon  ~=~\varepsilon \,.
 \label{projs3}
\end{equation}
This is still consistent with the projector (\ref{projs2}), allowing the addition of a set of M5' branes.  
One should also note that the sign in the momentum projector, $\Gamma^{01}$, is fixed implicitly by  the choice of frames and the sign of $P$ in  (\ref{11frames-null}).

The goal is, of course,  to solve
\begin{equation}
\delta \psi_\mu ~\equiv~ \nabla_\mu \, \epsilon ~+~ \coeff{1}{288}\,
\Big({\Gamma_\mu}^{\nu \rho \lambda \sigma} ~-~ 8\, \delta_\mu^\nu  \, 
\Gamma^{\rho \lambda \sigma} \Big)\, F_{\nu \rho \lambda \sigma} ~=~ 0 \,,
\label{11dgravvar}
\end{equation}
using the Ansatz for the metric and fluxes, subject to the foregoing projection conditions.

The dependence of the supersymmetries on the sphere directions is determined entirely by group theory.  With our choices of projectors and frames, the supersymmetries, $\varepsilon$, are independent of ${\varphi'}_j$, and independent of ${\varphi}_j$ for the Choice (ii) metric  (\ref{4metrics-1b}) with  (\ref{1forms}).  For the Choice (i) metric (\ref{4metrics-1c})  one must solve for Killing spinors on the $S^2$ and use the fact that $\chi$ is simply $\IR$ or $S^1$.

This yields:
\begin{equation}
\partial_{\theta} \varepsilon  ~=~\frac{1}{2}\, \Gamma^{35} \,\varepsilon \,, \qquad \partial_{\phi} \varepsilon  ~=~\frac{1}{2}\, \Big(\sin \theta \, \Gamma^{36} + \cos \theta \, \Gamma^{56} \Big) \,\varepsilon \,, \qquad \partial_{\chi} \varepsilon  ~=~ 0\,.
 \label{S2Ksp}
\end{equation}
The dependence of the spinors on $(\ncb, u, v, z)$ follows from the fact that $K^\mu \equiv \bar \varepsilon \Gamma^\mu \varepsilon$ is  the time-like Killing vector, $\frac{\partial}{\partial \nca}$.  This means that 
\begin{equation}
\varepsilon~=~ e^{\frac{1}{2} A_0} \, F^{-\frac{1}{2}} \, P^{-\frac{1}{4}}\, \varepsilon_0 \,,
\label{susyform1}
\end{equation}
where $\varepsilon_0$ is independent of all the coordinates for the metric (\ref{4metrics-1b}) with  (\ref{1forms}), or,   for the metric (\ref{4metrics-1c}), $\varepsilon_0$ only has the coordinate dependences implied by   (\ref{S2Ksp}).

\subsection{Outline of solving the BPS equations}
\label{ss:solvingsusies}

Since we know the coordinate dependences of $\varepsilon$, it is now straightforward to solve the BPS equations, (\ref{11dgravvar}), using the metric and flux Ansatz described above.   All but two of the $36$  flux functions are determined algebraically in terms of the metric functions and their first derivatives (most of the fluxes are identically zero).  One then finds simple sets of first-order equations that relate the metric functions to one another.  The computation proceeds much as in \cite{Bena:2023rzm}.  There is still some gauge freedom left in redefining the coordinates and this can be used to fix some of the metric functions completely.  

\subsubsection{The form of the metric and fluxes}
\label{ss:metflux}

After solving the supersymmetry transformations we find the metric reduces to the form:
\begin{equation}
\begin{aligned}
ds_{11}^2 ~=~  e^{2  A_0}\, \bigg[ &  d \ncb \, \bigg( P \,  d  \ncb    ~+~ 2\, \bigg( \frac{ d \nca}{F(\ncb) } ~+~ k \, \hat \sigma_3  \bigg)\bigg)~+~ e^{-3  A_0} \, (-\partial_z w )^{-\frac{1}{2}}\, ds_4^2 \\ & ~+~ e^{-3  A_0} \, (-\partial_z w )^{\frac{1}{2}}\,  {ds'}_4^2
 ~+~  (-\partial_z w ) \, \big( dz ~+~(\partial_z w )^{-1}\,  (\partial_u w ) \, d u \big)^2  \bigg]\,,
\end{aligned}
 \label{11metric-null-res}
\end{equation}
where the four-dimensional metrics are those of flat space.

 The BPS equations almost completely fix the relative scales, $A_4, A_5$ and $A_6$,  of the various pieces within $ds_{4}^2$ and ${ds'}_4^2$ so that these metrics are flat.  Note that, but for $P$ and $k$, the functions appearing in this metric are exactly the same as those of the substrate momentum-less solution reviewed in Section  \ref{sec:substrate}.
  
 There are some constants of integration that can be absorbed into coordinate re-definitions, but there is one constant that remains unfixed: one is allowed to have a constant re-scaling to the Hopf fiber of  $ds_{4}^2$ in Choice (ii).  We found that allowing this Hopf fiber to become squashed away from its round value led to singularities in the solution, and so we fixed the metric to that of a round $S^3$ in Choice (ii).  Thus,
\begin{equation}
 {ds'}_4^2  ~=~  dv^2  ~+~ \coeff{1}{4} \, v^2 \, \,({\sigma'}_1^2 +{\sigma'}_2^2+{\sigma'}_3^2\big)    \,,
 \label{4metrics-2as}
\end{equation}
and for Choice (i)  
\begin{equation}
ds_{4}^2 ~=~  du^2  ~+~ u^2 \,  \, \big(d\theta^2  + \sin^2\theta\, d\phi^2 \big)  ~+~ \, d\chi^2 \,,
 \label{4metrics-2bs}
\end{equation}
while for Choice (ii)
\begin{equation}
 ds_{4}^2 ~=~     du^2  ~+~ \coeff{1}{4} \, u^2 \, (\sigma_1^2 +\sigma_2^2+ \sigma_3^2 \big)  \,. \qquad 
 \label{4metrics-2c}
\end{equation}

The functions, $P, k, w$ and $A_0$ are, as yet, arbitrary functions of $(u,v,z)$.  The function, $F(\ncb)$, remains unconstrained.  

As indicated earlier, almost all the flux functions are determined in terms of metric functions.  Indeed, we find that the fluxes sourced by the M2 and M5 branes are related to metric functions exactly as they were in  \cite{Bena:2023rzm}.   The new non-zero  fluxes, again in frame indices, are:
\begin{equation}
\begin{aligned}
F_{0237} =&  -F_{1237} ~=~  b_1 \,, \qquad   F_{0347}=  -F_{1347} ~=~  b_2 \,, \qquad  F_{0247} =  -F_{1247} ~=~ \frac{1}{2}\,  \frac{e^{2 A_0}}{\sqrt{P}}\,   (\partial_v k) \,,   \\ F_{0256} = & -F_{1256} ~=~  -b_1~+~  \frac{1}{2}\,  \frac{e^{2 A_0}}{\sqrt{P}}\,   \bigg( (-\partial_z w )^{\frac{1}{2}}\,  \partial_u k+ \frac{\partial_u w}{(-\partial_z w )^{\frac{1}{2}}}\,  \partial_z k\bigg)  \,,
\end{aligned}
 \label{fluxfunctions}
\end{equation}
where $b_1, b_2$ are arbitrary functions of $(u,v,z)$. 
 These new components of the flux  satisfy  $F_{1b c d} = - F_{2 b c d}$, as one expects for null waves. 
 In Section \ref{ss:Match} we will show the charges of this solution are  those of the DBI solution reviewed in Section \ref{sec:Gluing}.

\subsubsection{The Bianchi Identities - I }
\label{ss:BianchiMet}

The heavy lifting in solving the BPS equations is to solve the Bianchi identities.  That is, we have made an Ansatz for all the possible terms in $F^{(4)}$, and determined these based on the supersymmetry, but one must now impose $dF^{(4)} =0$.  These equations fall into two parts: those of the \nBPS{4} substrate  and those for the new fluxes. Most significantly, the equations governing the  \nBPS{4} brane substrate completely decouple from the equations relating to the addition of the momentum wave.

Thus, the first set of Bianchi equations  turn out to be exactly the same as those for the \nBPS{4} background ``maze''  described in Section \ref{sec:substrate} and their solution proceeds in an identical manner to that described in \cite{Bena:2023rzm}.   That is, these equations  determine the functions $A_0$ and $w$ by solving (\ref{maze-eq}) and using (\ref{solfns}).

The remaining Bianchi identities determine the new fluxes and polarization vector, and depend on the functions $w$ and $A_0$.  However, the latter functions are now to be considered as part of the {\it known} background of the substrate branes.

\subsubsection{The Bianchi Identities - II }
\label{ss:BianchiFlux}

The Bianchi equations for the new fluxes  are rather non-trivial, but they are {\it linear} in the fluxes.  We consider  Choice (i) and Choice (ii) separately, as the equations are slightly different. 

For Choice (i), one of the Bianchi identities can be written as:  
\begin{equation}
\begin{aligned}
& \partial_u\Big[ \, u^2  \, e^{-{\frac{1}{2} A_0}}(-\partial_z w)^{-\frac{1}{4}} \, \sqrt{P} \,b_2 \, \Big]     \\
& = \partial_v\bigg [ \frac{ u^2 }{(\partial_z w)}\,\bigg( \frac{1}{2} \,\partial_z k -  e^{ A_0} (\partial_u w)  \Big( \sqrt{P}\, b_1- \frac{1}{2} e^{ 2A_0} \big((-\partial_z w)^{\frac{1}{2}}\, \partial_u k + (-\partial_z w)^{-\frac{1}{2}}\,  (\partial_u w)  \, \partial_z k \, \big)\Big)\bigg)\, \bigg]  \,.
 \end{aligned}
 \label{intBianchi1}
\end{equation}
This can be satisfied by introducing a pre-potential, $q$, defined by: 
\begin{equation}
\begin{aligned}
\partial_v q ~=~ &   \, u^2 \,e^{-{\frac{1}{2} A_0}}(-\partial_z w)^{-\frac{1}{4}} \,  \sqrt{P} \, b_2  \,,   \\
\partial_u q ~=~ &  \frac{ u^2 }{(\partial_z w)}\,\bigg( \frac{1}{2} \,\partial_z k -  e^{ A_0} (\partial_u w)  \Big( \sqrt{P}\, b_1- \frac{1}{2} e^{ 2A_0} \big((-\partial_z w)^{\frac{1}{2}}\, \partial_u k + (-\partial_z w)^{-\frac{1}{2}}\,  (\partial_u w)  \, \partial_z k \, \big)\Big)\bigg)  \,.
 \end{aligned}
 \label{intBianchi2}
\end{equation}
Using these identities to replace $b_1$ and $b_2$ in the remaining Bianchi identities leads to two more equations, one of which is relatively simple, and can be solved by introducing another pre-potential, $p$, defined by: 
\begin{equation}
\partial_u p ~=~ u^2 \, k ~+~ 2\,(\partial_u w)\, q  \,, \qquad   \partial_v p ~=~   2\,(\partial_z w) \, q   \,.
 \label{intBianchi3}
\end{equation}
One then finds that this pre-potential, $p$ also solves the remaining Bianchi identity.

Note that the polarization vector of the null momentum wave is given by:
\begin{equation}
k ~=~\frac{1}{u^2} \, \bigg(\partial_u p~-~ \frac{(\partial_u w)}{(\partial_z w)}\, \partial_v  p \bigg)    \,.
 \label{kform1}
\end{equation}

The analysis for Choice (ii) is almost identical, except that the pre-potentials are defined by
\begin{equation}
\begin{aligned}
\partial_v q ~=~ &  u^4\, e^{-{\frac{1}{2} A_0}}(-\partial_z w)^{-\frac{1}{4}} \,  \sqrt{P} \, b_2  \,,  \\
\partial_u q ~=~ & \frac{ u^3 }{(\partial_z w)}\,\bigg( \partial_z k -  e^{ A_0} (\partial_u w)  \Big( u\, \sqrt{P}\, b_1-   e^{ 2A_0} \big((-\partial_z w)^{\frac{1}{2}}\, \partial_u k + (-\partial_z w)^{-\frac{1}{2}}\,  (\partial_u w)  \, \partial_z k \, \big)\Big)\bigg) \,,
 \end{aligned}
 \label{intBianchi4}
\end{equation}
and
\begin{equation}
\partial_u p ~=~2 \big( u^3 \, k ~+~ (\partial_u w)\, q  \big) \,, \qquad   \partial_v p ~=~   2\,(\partial_z w) \, q   \,,
 \label{intBianchi5}
\end{equation}
and hence one has
\begin{equation}
k ~=~ \frac{1}{2 u^3} \, \bigg( \partial_u p~-~ \frac{(\partial_u w)}{(\partial_z w)}\, \partial_v  p \bigg)    \,.
 \label{kform2}
\end{equation}
As with Choice (i), one finds that using the pre-potential, $p$  solves the remaining Bianchi identities.

We have thus reduced the solving of the fluxes and polarization vector, $k$,  to finding a single, undetermined function, $p$.   Indeed, the rest of the solution is contained by two undetermined functions:  The momentum density, $P$,  and the pre-potential $p$, both of which will be governed by the equations of motion.

\subsubsection{The gauge potential}
\label{ss:Cres}

Given that we have solved the Bianchi identities, we can now integrate the flux Ansatz to obtain an expression for the gauge potential.  
For Choice (i) we find:
\begin{equation}
\begin{aligned}
 C^{(3)} ~=~ &  -  e^0 \wedge e^1 \wedge e^2 \\ & - \bigg(\frac{\partial_u w}{\partial_z w}\bigg)\,  u^n \sin \theta  \, d\theta \wedge d\phi  \wedge d\chi
 + \frac{1}{8} \, (\partial_v w)\,v^3 \sin {\varphi'}_1 \, d{\varphi'}_1 \wedge d{\varphi'}_2  \wedge d{\varphi'}_3 \\
 & +  \frac{1}  {u^\ell \sqrt{P}\, (-\partial_z w)^{\frac{1}{2}}} \, \big(\partial_z p \big)  \, (e^1-e^0)\wedge \big( e^3\wedge e^7~-~ e^5\wedge e^6 \big)   \,, \label{C3res1-1}
 \end{aligned}
\end{equation}
with $\ell =2, n=2$, while for Choice (ii) we find
\begin{equation}
\begin{aligned}
 C^{(3)} ~=~ &  -  e^0 \wedge e^1 \wedge e^2 \\ & + \frac{1}{8} \,\bigg(\frac{\partial_u w}{\partial_z w}\bigg)\,  u^n \sin \varphi_1 \, d\varphi_1 \wedge d\varphi_2  \wedge d\varphi_3
 + \frac{1}{8} \, (\partial_v w)\,v^3 \sin {\varphi'}_1 \, d{\varphi'}_1 \wedge d{\varphi'}_2  \wedge d{\varphi'}_3 \\
 & +  \frac{1}  {u^\ell \sqrt{P}\, (-\partial_z w)^{\frac{1}{2}}} \, \big(\partial_z p \big)  \, (e^1-e^0)\wedge \big( e^3\wedge e^7~-~ e^5\wedge e^6 \big)   \,, \label{C3res1-2}
 \end{aligned}
\end{equation}
with $\ell =4, n=3$.

Naively, the first three terms in these expressions are exactly what one expects for the  fluxes of the \nBPS{4} background maze solution  \cite{Bena:2023rzm} discussed in Section \ref{sec:substrate}.  
 However, this perspective is a little oversimplified because one should remember that the frames $e^0$ and $e^1$ now involve a momentum density and a polarization vector, and are therefore significantly more complicated than those of   \cite{Bena:2023rzm}.

\subsubsection{Matching to the Born-Infeld construction}
\label{ss:Match}

It is interesting to try to connect the features of the supergravity solution to those of the DBI description of the D4-D2 momentum wave  \cite{Bena:2023fjx} reviewed in Section \ref{sec:Gluing}. 

To do this, one first has to re-define the null coordinates, such that the $d\nca~d\ncb$ term in the metric is $\xi$ independent. This is done by introducing a new null coordinate, $\eta$, such that 
\begin{equation}
d \eta = {d \xi \over F(\xi)} \equiv   {d \xi \over f(\eta)} \,.
\label{etadef}
\end{equation}
The function $f(\eta)$ is defined implicitly above and, since  $F(\xi)$ is an arbitrary function, one can also consider  $f(\eta)$ as the defining arbitrary function of our solution, which is now written as
\begin{equation}
\begin{aligned}
ds_{11}^2 ~=~  e^{2  A_0}\,  &  d \eta \,\big[ 2\,   d \nca      ~+~  2 k f(\eta) \, d \chi ~+~ P f(\eta)^2 \,  d  \eta \big] ~+~ e^{-  A_0} \, (-\partial_z w )^{-\frac{1}{2}}\, ds_4^2 \\ & ~+~ e^{-  A_0} \, (-\partial_z w )^{\frac{1}{2}}\,  {ds'}_4^2
 ~+~  e^{2  A_0}\,   (-\partial_z w ) \, \big( dz ~+~(\partial_z w )^{-1}\,  (\partial_u w )~du \big)^2 \,,
\end{aligned}
 \label{11metric-null-eta}
\end{equation}
where we have explicitly used the Choice (i) metric, which is related to the brane construction in Section \ref{sec:Gluing}. The metric already allows us to see one of the components of the glue, corresponding to momentum along the $\chi$ direction (depicted on the right side of the triangle in Figure \ref{gluing}(b)). As expected from the DBI construction, this glue charge is parameterized by the arbitrary function $f(\eta)$.

To see the other components of the glue, it is best to explicitly expand the vielbeins in the Choice (1) gauge potential \eqref{C3res1-1} and use the null coordinate $\eta$ introduced in \eqref{etadef} 
\begin{equation}
\begin{aligned}
 C^{(3)} ~=~ &   e^{3  A_0} \, (-\partial_z w )^{1\over 2} \, d \eta \wedge \big(d \zeta  + k f(\eta) \,d\chi\big)  \wedge \big( dz ~+~(\partial_z w )^{-1}\,  (\partial_u w )~du \big) \\ 
 & -  \,\bigg(\frac{\partial_u w}{\partial_z w}\bigg)\,  u^2 \sin \theta  \, d\theta \wedge d\phi  \wedge d\chi
 + \frac{1}{8} \, (\partial_v w)\,v^3 \sin {\varphi'}_1 \, d{\varphi'}_1 \wedge d{\varphi'}_2  \wedge d{\varphi'}_3 \\
 & + \frac{\big(\partial_z p \big)}  { u^2  } \,  f(\eta) \, d \eta \wedge \left(du \wedge d \chi - u^2 \sin\theta \, d \theta \wedge d \phi \right) 
   \,, \label{C3expanded}
 \end{aligned}
\end{equation}
%
 %
 %

On can see from the first line of this potential that the solution also has the other glue charge depicted on the right side of the triangle in Figure \ref{gluing}(b), corresponding to M2 branes extended along the $\chi$ and $z$ directions. 

The last line of this expression also makes explicit the equality of the M2$_{\theta,\phi}$ and M2$_{u,\chi}$ glue charges, shown on the top line of triangle in Figure \ref{gluing}(b). These charges are the M-theory uplift of the charges \eqref{M2-glue} of the DBI construction in Section \ref{sec:Gluing}, and their equality is also a consequence of the DBI construction.

\subsection{The equations of motion}
\label{ss:eoms}

Having solved the BPS equations, one must still check the equations of motion.  This will determine the remaining functions, $p$ and $P$.

It is useful to compute the Laplacian, $\hat {\cal  L}$, for the  substrate metric (\ref{11metric}) acting on a  function  $H$ that only depends on $(u,v,z)$.   Using symmetries we have imposed, and the equations for $w$ and $A_0$,  one can simplify the Laplacian  to obtain the following operator: 
\begin{equation}
\begin{aligned}
{\cal L} ( H) ~\equiv~ &e^{- A_0}\,(-\partial_z w )^{-\frac{1}{2}} \,   \hat {\cal L} ( H)  \\
 ~=~   &   \bigg[  \frac{1}{u^n}\, \partial_u \big( u^n  \partial_u H \big) ~+~  \frac{1}{ (- \partial_z w)}\,  \frac{1}{v^3} \partial_v \big( v^3  \partial_v H \big) ~+~2\, \frac{( \partial_u w)}{ (- \partial_z w)} \, \partial_u \partial_z H     \\
  & ~+~  \Big((-\partial_z  w)^{-\frac{3}{2}} \,e^{-3  A_0}~+~  (-\partial_z  w)^{-2} \,  (\partial_u w)^2\big)\Big)\, \partial_z^2 H \bigg] \,,
\end{aligned}
 \label{Lap}
\end{equation}
with $n =2$ or $n =3$ for choices (i) or (ii), respectively.  It is interesting to note that one could also have replaced $\hat {\cal  L}$ by the Laplacian for the final metric (\ref{11metric-null-res}) because {\it on functions of $(u,v,z)$ alone}, these two Laplacians agree. Here, however, we wish to emphasize that ${\cal L}$ is a Laplacian operator on a known substrate background that does not depend upon the functions we are trying to determine. 

The Maxwell equations actually give rise to two rather different-looking  {\it third-order} linear equations for $p$.  Fortunately, these two equations are compatible, and can be solved by requiring $p$ to be the solution of a {\it single, second order, linear equation}:
\begin{equation}
{\cal L} \bigg( \frac{p}{u^\ell}\bigg) ~-~ \frac{2\,m }{u^2}\, \frac{p}{u^\ell}~=~  0 \,,
 \label{peqn}
\end{equation}
with $\ell =2, m=1$ or $\ell =4, m=4$ for choices (i) or (ii), respectively.  To be more specific, the two seemingly-independent Maxwell equations are actually combinations of the differential equation  (\ref{peqn}) and either the $u$-derivative, or the $z$-derivative, of this differential equation.

The Einstein equations  produce  second-order differential equations for $k$ and $P$.  The former is identically satisfied if one uses (\ref{kform1}), or  (\ref{kform2}), to rewrite $k$ in terms of $p$,  and then employs  (\ref{peqn}) (or the combinations of  (\ref{peqn}) that arise in the Maxwell equations).   

The equation for $P$ can also be written as:
\begin{equation}
{\cal L} \big(P\big)  ~=~  s_{(x)} \,,
 \label{Peqn}
\end{equation}
where, for the two choices, one has:
\begin{equation}
\begin{aligned}
s_{({\rm i})} = - 4\,e^{- A_0}\,(-\partial_z w )^{-\frac{1}{2}} \,\bigg[& 2\,\Big(\big(\sqrt{P} \, b_1\big)^2 +\big(\sqrt{P} \, b_2\big)^2  \Big) \\
&- e^{2 A_0}\, \Big(\sqrt{P} \, b_1\Big) \big((-\partial_z w )^{\frac{1}{2}} \, \partial_u k  ~+~ (\partial_u w) \, (-\partial_z w )^{-\frac{1}{2}}\, \partial_z k \big)\bigg] \,, \\
s_{({\rm ii})}  = - 8\,e^{- A_0}\,(-\partial_z w )^{-\frac{1}{2}} \,\bigg[&  \big(\sqrt{P} \, b_2\big)^2 
~+~  \bigg(\big(\sqrt{P} \, b_1\big) ~-~ \frac{2\, e^{2 A_0}}{u^2}\,(-\partial_z w )^{\frac{1}{2}}\,k\, \bigg)  \\ &  \times \bigg(\big(\sqrt{P} \, b_1\big) ~-~ \frac{e^{2 A_0}}{u} \,\Big((-\partial_z w )^{\frac{1}{2}}\,\partial_u k~+~ (-\partial_z w )^{-\frac{1}{2}}\,(\partial_u w)\,\partial_z k\,\Big) \bigg)\bigg]\,. 
\end{aligned}
 \label{sources}
\end{equation}
The important point here is that $\sqrt{P} \, b_{1,2}$ can be eliminated via  (\ref{intBianchi2}) and (\ref{intBianchi3}) or (\ref{intBianchi4}) and (\ref{intBianchi5}),  to obtain sources, $s_{(x)}$, that are completely independent of $P$, and only depend on the known fields, $w$, $A_0$, and $p$.   This means that the equation for $P$,  (\ref{Peqn}), is {\it linear}, and sourced by the background fluxes and metric components that have already been determined.


\subsection{An interesting footnote}
\label{ss:inhomogeneous}

The equation for $P$, (\ref{Peqn}), is inhomogeneous and the sources, (\ref{sources}), are very complicated once they are expanded using  (\ref{intBianchi2}) and (\ref{intBianchi3}), or (\ref{intBianchi4}) and (\ref{intBianchi5}).  However, motivated by similar equations in other microstate geometries, one can make a simple guess for part of the required ``particular solution.'' We find:
\begin{equation}
\begin{aligned}
{\cal L} \bigg(&\frac{(\partial_z p)^2}{u^4 (\partial_z w)}\bigg)  ~-~  s_{({\rm i})} \\
&=~ \frac{2}{u^6 (\partial_z w)} \, \Big(6 (\partial_z p)^2 - 4 u (\partial_z p)  (\partial_u \partial_z p) + 2 u (\partial_z^2 p)  (\partial_u  p)~+~ u^2 \big(  (\partial_u \partial_z p)^2  - (\partial_u^2  p) (\partial_z^2  p)\big)   \Big)  \,,  \\ 
{\cal L} \bigg(&\frac{(\partial_z p)^2}{u^8 (\partial_z w)}\bigg)  ~-~  s_{({\rm ii})} \\
&=~ \frac{1}{u^{10} (\partial_z w)} \, \Big(24 (\partial_z p)^2 - 8 u (\partial_z p)  (\partial_u \partial_z p) + 5 u (\partial_z^2 p)  (\partial_u  p)~+~ u^2 \big(  (\partial_u \partial_z p)^2  - (\partial_u^2  p) (\partial_z^2  p)\big)   \Big)  \,. \\ 
\end{aligned}
 \label{inhomsols}
\end{equation}
These partial solutions to the inhomogeneous terms represent a very substantial simplification of the source terms, but we have not found a simple expression for a particular solution that generates the right-hand sides of (\ref{inhomsols}).

\subsection{Summary of the solution}
\label{ss:summary}

The \nBPS{8} solution carrying momentum waves starts from a \nBPS{4} M2-M5 substrate whose metric is given by (\ref{11metric}) and fluxes are given by  (\ref{C3gen}).  The unknown functions, $w$ and $A_0$ are obtained by solving (\ref{maze-eq})  and using  (\ref{solfns}).

Imposing  symmetries as described in Section \ref{ss:metricfluxAnsatz}, the metric with momentum waves is given by (\ref{11metric-null-res}) and the frames are defined in (\ref{11frames-null}). The fluxes are now determined in terms of a pre-potential, $p$, via (\ref{C3res1-1}) and (\ref{C3res1-2}) and the polarization function, $k$, is determined in terms of $p$ through (\ref{kform1}), or (\ref{kform2}), depending upon the imposed symmetries. 

The pre-potential, $p$, is determined by a modified Laplace equation,  (\ref{peqn}) with operator (\ref{Lap})  defined by the  substrate metric (\ref{11metric}).  The momentum density, $P$, is fixed by a Poisson equation, (\ref{Peqn}), also with  operator (\ref{Lap}), but now  with sources, (\ref{sources}).  The crucial fact is that this last set of equations in actually {\it linear}.  Indeed, the only non-linear equation to be solved is (\ref{maze-eq}) which defines the \nBPS{4}  substrate.

The wave profile, $F(\ncb)$, is a freely choosable (arbitrary) function.

The sources of the Poisson equation for the momentum density, $P$, are quadratic in terms that define the momentum flux.  This is a vestige of the Chern-Simons interaction in the equation for $F^{(4)}$. A partial particular solution to the inhomogeneous equation can be obtained from the squares of derivatives of the pre-potential as demonstrated in  (\ref{inhomsols}).  

\subsection{A conjecture about multiple momentum waves.}
\label{ss:conjecture}

The fact that the full momentum-carrying solution is constructed on top of a substrate by null waves in the ``glue'' fields which can be added in a linear procedure suggests a very obvious generalization of our solution. 

First, note that the substrate in equations \eqref{11frames},\eqref{C3gen} does not need to have any spherical symmetry, and can describe in principle a multitude of M2 brane strips stretching between M5 branes. Our linear system also suggests an obvious  covariantization.

The fundamental object will be an anti-self-dual form on the $\IR^4$ wrapped by the M5 branes. In the third line of equation \eqref{C3res1-1}  this appears as 
\begin{equation}
 {\cal{P}}\equiv
 \, \big(\partial_z p \big)  \, \big( e^3\wedge e^7~-~ e^5\wedge e^6 \big)   \,,
\label{Pform}
\end{equation}
and one can see from \eqref{C3expanded} that it can be multiplied by an arbitrary function of $\eta$. The homogeneous equation for $p$, \eqref{peqn} will emerge from the anti-self-duality.
Equation \eqref{intBianchi3} shows that the one-form $k$ will emerge as a combination of the divergence of $p$ and the inner product of $\grad w$ and $p$. The function $q$ and hence the fluxes $\sqrt{P} b_1$ and $ \sqrt{P} b_2$ are also 2-forms, which we will denote schematically as ``$\cal{B}$''. All these fields are part of the ``glue'' and is parameterized again by an arbitrary function of $\eta$. Finally, equations \eqref{Peqn} and \eqref{sources} show that the momentum density is sourced by terms of the form $*_4 (\cal{B}\wedge\cal{B})$,  $*_4 (k \wedge \grad w \wedge \cal{B})$ and  $*_4 (d k \wedge\cal{B})$.

It is reasonable to assume that one can source the anti-self-dual form $p$ independently on each strip, and that the corresponding waves will be parameterized by different arbitrary functions of $\eta$.  Hence, the metric will become 
\begin{equation}
\begin{aligned}
ds_{11}^2 ~=~  e^{2  A_0}\,  &  d \eta \,\big[ 2\,   d \nca      ~+~  2 f^i(\eta) \vec{k_i} \cdot d \vec{u} ~+~ P_{ij} f^i(\eta) f^j(\eta) \,  d  \eta \big] ~+~ e^{-  A_0} \, (-\partial_z w )^{-\frac{1}{2}}\, ds_4^2 \\ & ~+~ e^{-  A_0} \, (-\partial_z w )^{\frac{1}{2}}\,  {ds'}_4^2
 ~+~  e^{2  A_0}\,   (-\partial_z w ) \, \Big( dz ~+~(\partial_z w )^{-1}\,  \big (\vec \nabla_{\vec u} \, w \big)  \cdot  d \vec u \Big) \,,
 \end{aligned}
 \label{11metric-null-eta-new}
\end{equation}
where we have used the label $i$ to enumerate different strips and their sources. Since $P$ is sourced quadratically, it will carry two of these enumeration indices.  Furthermore, the potential will be
\begin{equation}
\begin{aligned}
 C^{(3)} ~=~ &   e^{3  A_0} \, (-\partial_z w )^{1\over 2} \, d \eta \wedge \big(d \zeta  + 
 f^i(\eta) \vec{k_i} \cdot d \vec{u}\,\big)  \wedge \Big( dz ~+~(\partial_z w )^{-1}\,  \big (\vec \nabla_{\vec u} \, w \big)  \cdot  d \vec u \Big) \\ 
 + &\frac{1}{3!}\, \epsilon_{ijk\ell} \,  \big((\partial_z w )^{-1}\, (\partial_{u_\ell} w) \,  du^i \wedge du^j \wedge du^k ~-~ (\partial_{v_\ell} w)  \, dv^i \wedge dv^j \wedge dv^k  \big)  + \,  f^i(\eta) \, d \eta \wedge {\cal{P}}_i
   \,, \label{C3expanded-new}
 \end{aligned}
\end{equation}
where $ {\cal{P}}_i$ is the suitable generalization of \eqref{Pform}. 

Obviously, there is much here that needs to be verified, but we are optimistic based on the linearity and our experience with superstrata.

\section{Final comments}
\label{sec:Conc}

The construction of Microstate Geometries for black holes started almost 20 years ago with \cite{Bena:2006kb,Bena:2007kg}.  This work was motivated by the desire to extend Mathur's remarkable fuzzball program from two-charge solutions, to the ``three-charge problem,'' for which the corresponding black holes have macroscopic horizons.  Ironically, the expectation of one of the authors was that the BPS equations would be hopelessly non-linear because having three independent sets of charges and magnetic fluxes would activate the Chern-Simons interaction, and this would make the Maxwell equation non-linear.  To our very great surprise, the system of BPS equations governing the fluxes turned out to be linear \cite{Bena:2004de}:  the non-linearities were confined to the source terms that were quadratic in known solutions to other linear equations. 

This result opened up the analysis of the phase space of five-dimensional microstate geometries.  While there were very large numbers of solutions \cite{Bena:2006is,Bena:2007kg,Bena:2007qc,  Bianchi:2016bgx, Bianchi:2017bxl, Heidmann:2017cxt, Bena:2017fvm, Avila:2017pwi, Tyukov:2018ypq, Warner:2019jll}, and some of them had the mass gap of the typical states of the dual CFT, such geometries could only account for a tiny fraction of the black-hole microstates.  These five-dimensional microstate geometries were only accessing CFT states that had a $U(1) \times U(1)$ isometry.  It became imperative to break these symmetries and add more degrees of freedom.  The  obvious extension was to go to six dimensions and incorporate and generalize the supertubes that had been such an integral part of the two-charge fuzzballs.  In yet another irony, the other author of the Microstate Geometry program was deeply skeptical that the equations \cite{Gutowski:2003rg,Cariglia:2004kk} governing six-dimensional microstate geometries  could also be linear. But they were \cite{Bena:2011dd}!

More precisely,  in both five and six-dimensions, the equations governing the  substrate geometries were non-linear (Monge-Amp\`ere) equations governing either hyper-K\"ahler, or almost hyper-K\"ahler substrate geometries \cite{Gauntlett:2002nw,Gutowski:2003rg}.  Adding momentum, and the fluxes to carry the momentum, is entirely governed by a linear system of equations \cite{Bena:2004de,Bena:2011dd,Ceplak:2022wri}.  From a thorough analysis of the CFT duals of these six-dimensional geometries, it became clear that  the substrate geometry was determining the CFT, or IR ground state and the linearly determined fluxes and momentum carriers were dual to families of  CFT  excitations of these ground states \cite{Bena:2015bea,Giusto:2015dfa}.

 The linearity of the supergravity solutions was  essential to both the development of the holographic dictionary and the analysis of the CFT excitations.

Over the last few years,  the six-dimensional system has been extensively mapped out with precision holography ~\cite{%
Galliani:2016cai,
Bombini:2017sge,
Bombini:2019vnc,
Tian:2019ash,
Tormo:2019yus,
Giusto:2019qig,
Bena:2019azk,
Giusto:2019pxc,
Hulik:2019pwr,
Giusto:2020mup,
Ceplak:2021wzz,
Shigemori:2021pir,
Rawash:2021pik}.
 We know the strengths and limitations of this system and we know exactly what states it can capture and what it is missing.  There are a vast number of states accessible to the six-dimensional system, but their entropy grows, at most, as $\sqrt{Q_1 Q_5 }\sqrt[4]{Q_p}$, \cite{Shigemori:2019orj,Mayerson:2020acj} which is parametrically short of the black-hole entropy,  $\sqrt{Q_1 Q_5 Q_p}$.  The shortfall comes from the fact that six-dimensional supergravity cannot resolve the brane fractionation that is essential to accessing the twisted sector states.

This has led to a new thrust in which one tries to resolve brane fractionation using supergravity in ten or eleven dimensions. The simple idea is that since there are a truly vast number of microstates, then there should also be an exceptional  number of coherent expressions of those microstates that will be visible in supergravity.  Hence brane fractionation should have a supergravity avatar.  In much the same way that the analysis of supersymmetric brane configurations \cite{Bena:2011uw} led to  {\it superstrata} in six-dimensional supergravity \cite{Bena:2015bea,Bena:2016ypk,Bena:2017xbt}, a similar analysis of super-mazes and themelia   \cite{Bena:2022wpl, Bena:2022fzf, Bena:2023fjx}  led to  the work presented here, and once again we seem to be finding the same miracles.

The \nBPS{4} substrate geometries  are determined by complicated, non-linear equations.  However, the momentum excitations, and the fluxes that carry them, appear to be governed by linear systems.  This strongly suggests that the substrate geometries (and their non-linear equations) determine the particular twisted, or fractionated, sector of the dual field theory, and once again the momentum excitations, and the states that carry them, are determined by a linear system.    What remains to be done is much like the story of superstrata:  we need to find the most general families of momentum excitations, and geometric transitions of the super-maze geometries and map out the states accessible to supergravity.  The linearity we have discovered here and the discussion in Section \ref{ss:conjecture} suggests that this is going to be a feasible undertaking.

This would mean  that supergravity can access  the twisted sectors of the CFT {\it and} enable one to fully analyze the phase space of the momentum excitations within those twisted sectors.  As a result,  supergravity could be used to sample all the essential sectors of the underlying CFT and see the details of the states that make up the black-hole microstructure.  The entropy of such microstate geometries  should grow as $Q^{3/2}$. 

The linear description of this phase space will not only prove critical to counting the microstates, but it may well enable the development of precision holography of those states.   

In retrospect, we now believe we understand the ``why'' of all the linear systems emerging from microstate geometries, and this is the heart of the extended themelion conjecture: the linear systems are a feature of the ``glue'' that welds the momentum to the branes to create an object that has sixteen local supersymmetries.  The  sixteen local supersymmetries are responsible for the local phenomenon of linearity of the equations that govern the excitations.

\vspace{1em}\noindent {\bf Acknowledgements:} \\
We would like to thank Costas Bachas, Nejc \v Ceplak, Soumangsu Chakraborty, Eric D'Hoker, Shaun Hampton, Yixuan Li and  Emil Martinec for interesting discussions.
The work of IB, AH and NPW was supported in part by the ERC Grant 787320 - QBH Structure. The work of IB was also supported in part by the ERC Grant 772408 - Stringlandscape and by the NSF grant PHY-2309135 to the Kavli Institute for Theoretical Physics (KITP).
 The work of AH was also supported in part by a grant from the Swiss National Science Foundation, as well as via the NCCR SwissMAP. The work of DT was supported in part by the Onassis Foundation - Scholarship ID: F ZN 078-2/2023-2024 and by an educational grant from the A. G. Leventis Foundation. The work of NPW was also supported in part by the DOE grant DE-SC0011687.
\newpage



\newpage

\begin{adjustwidth}{-1mm}{-1mm} 

\bibliographystyle{utphys}      

\bibliography{references}       

\end{adjustwidth}


\end{document}